\documentclass[aps,pra,twocolumn,showpacs,preprintnumbers,amsmath,amssymb]{revtex4}

\usepackage{graphicx}
\usepackage{dcolumn}
\usepackage{color}
\usepackage{bm}
\usepackage{amsmath}

\usepackage{dcolumn}

\begin{document}

\title{Experimental tomography of NOON states with large photon numbers}
\author{Y. Israel}

\author{I. Afek}%
\author{S. Rosen}
\author{O. Ambar}
\author{Y. Silberberg}%

\affiliation{%
Department of Physics of Complex Systems, Weizmann Institute of
Science, Rehovot 76100, Israel }
\date{\today}
\begin{abstract}
We have performed experimental quantum state tomography of NOON states with up to four photons. The measured states are generated by mixing a classical coherent state with spontaneous parametric down-conversion. We show that this method produces states which exhibit a high fidelity with ideal NOON states. The fidelity is limited by the overlap of the two-photon down-conversion state with any two photons originating from the coherent state, for which we introduce and measure a figure of merit. A second limitation on the fidelity set by the total setup transmission is discussed. We also apply the same tomography procedure for characterizing correlated photon hole states.
\end{abstract}
\maketitle

\begin{figure}[hhh!!]
\includegraphics{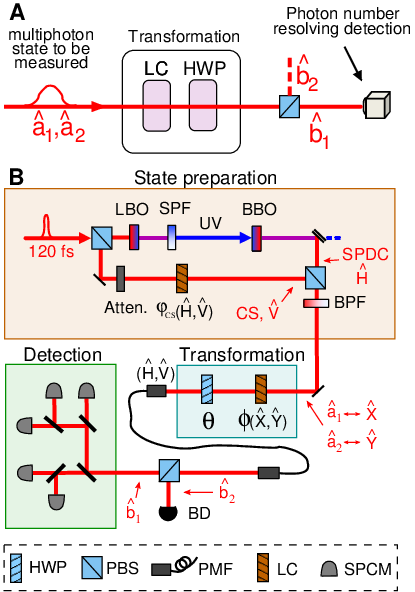}
\caption{\label{fig:setup} (Color online) Experimental setup for quantum state tomography of a multiphoton state in two modes. \textbf{(a)} Schematic of the setup depicting a prepared beam forming a multiphoton quantum state in two orthogonal polarization modes $\hat{a}_1$ and $\hat{a}_2$, undergoing a linear state transformation introduced by a half wave plate (HWP) and a liquid crystal (LC). The transformed state is photon-number resolved, in one of its polarization modes $\hat{b}_1$. \textbf{(b)} Detailed layout of the setup. A pulsed Ti:Sapphire oscillator with $120$ fs FWHM pulse width and $80$ MHz repetition rate is doubled using a $2.74$mm LBO crystal to obtain $404$nm  ultra-violet pulses with maximum power of $225$mW. These pulses then pump collinear degenerate type-I SPDC at a wavelength of $808$nm  using a $1.78$mm long BBO crystal. The SPDC \mbox{($\hat{H}$ polarization)} is spatially and temporally overlapped with a CS ($\hat{V}$ polarization) using a polarizing beam-splitter cube (PBS). A thermally induced drift in the relative phase, $\varphi_{cs}$, between $\hat{H}$ and $\hat{V}$ polarizations is corrected every few minutes using a LC phase retarder. The coherent light amplitude is adjusted using a variable attenuator. Modes $\hat{a}_1$,$\hat{a}_2$ are realized collinearly at the $\pm 45^{\circ}$ $\hat{X}, \hat{Y}$ polarizations respectively. These modes then pass through a LC aligned with the $\hat{X}$, and $\hat{Y}$ polarizations introducing a relative phase shift $\phi$ between them, and a HWP rotated by $\theta$ in the $\hat{H}$ and $\hat{V}$ plane. Modes $\hat{b}_1$, $\hat{b}_2$ are oriented at polarization axes $\hat{H}$ and $\hat{V}$ respectively. The spatial and spectral modes are matched using a $2$m long PMF and a $3$nm FWHM band-pass (BPF) filter centered at $808$nm. Photon number resolving detection is performed using multiple single photon counting modules (Perkin Elmer). Additional components: Dichroic mirrors, short-pass filter (SPF).}
\end{figure}
\section{Introduction}

Quantum state tomography (QST) is a well known method for full characterization of experimentally formed quantum states. It is common in experimental quantum optics to generate multiphoton quantum states in which each spatial mode is occupied by exactly one photon \cite{TwoQBits99PRL,WIECZOREK09PRL}. More generally, one could consider multiphoton states in which each optical mode may accommodate more than one photon \cite{2PhNOONdAngelo01,MITCHELL04NAT,SHALM09NAT,kim2009threeNOON,AFEKSCIENCE,MOONStates2011,PRL04EIS,Bogdanov04PRL,EPHFRANSONPRL06,CPH10PRL}. 'NOON states' which are N photon states in which all the photons occupy either one mode or the other are an example of such states. The name 'NOON' stems from their mathematical Fock state notation \cite{IntroQuantOPt,CONTEMPPHYS08DOWLING}
\begin{align}
| NOON \rangle_{A,B} \equiv \frac{1}{\sqrt{2}}(| N , 0 \rangle_{A,B} + | 0 , N \rangle_{A,B})
\end{align}
 where A and B denote two optical modes.  It was recently shown that NOON states can be generated with large photon numbers (high-NOON)\cite{AFEKSCIENCE}, by mixing a classical coherent state (CS) and quantum spontaneous parametric down-converted (SPDC) light on a standard beam-splitter \cite{HofmannPRA2007,Pezze2008PRL}. The resulting states were measured by showing super-resolution in the $N$-fold coincidence measurement. However, full characterization of the resulting states requires QST. In this paper we experimentally implement a method for performing QST measurements on high photon number states in two modes. The method is based on a convex optimization algorithm for the maximum-likelihood function. In addition, we apply the method to reconstruct the density matrices for another class of states, the correlated photon holes (CPH) \cite{EPHFRANSONPRL06,CPH10PRL}.

\section{Mathematical analysis}

The density matrix of a general mixed-state of $N$ indistinguishable photons in two modes, assumed here to be the two orthogonal polarization modes of one spatial mode, has $( N + 1 )^2$ elements that correspond to all $N$th-order coherences.
These $N$th-order coherences are expressed by \cite{glauber1963coherences}

\begin{align}
\rho_{m n} =  \langle\hat{a}_1^{\dagger n} \hat{a}_2^{\dagger N-n} \hat{a}_1^{m} \hat{a}_2^{N-m}\rangle
\end{align}
where $\hat{a}_1$ and $\hat{a}_2$ are the two operators of the two orthogonal modes of the electric field.
It was recently shown that by measuring only $N$-fold coincidence, it is possible to determine all $N$th order coherences for arbitrary $N$, using $SU(2)$ transformations \cite{Coherences10PRA}.
Using Jones calculus notation, we can parameterize this transformation, using two angles ($\theta$ and $\phi$), as
\begin{align}
\left(\begin{array}{c}
\hat{b}_1 \\
\hat{b}_2
    \end{array}\right)
    = U( \theta , \phi )
    \left(\begin{array}{c}
\hat{a}_1 \\
\hat{a}_2
    \end{array}\right) \label{eq:coord-trans}
\end{align}
where $\hat{b}_1$ and $\hat{b}_2$ are the two orthogonal operators of the electric field, representing the state following the linear transformation $U( \theta , \phi )$.
 This transformation can be implemented in many equivalent ways \cite{Coherences10PRA,SHALM09NAT}, one of which includes using a liquid crystal (LC) as a retarder of phase $\phi$, followed by a half wave plate (HWP) with an orientation angle of $\theta+\pi/4$ with respect to the LC,
\begin{align}
\nonumber  &U( \theta , \phi ) = \\
&\left(\begin{array}{lc}
\cos\theta & \sin\theta \\ \sin\theta & -\cos\theta
\end{array}\right)
\times
\frac{1}{\sqrt{2}} \left(\begin{array}{cc}
1 & -1 \\ 1 & 1
\end{array}\right)
\times
\left(\begin{array}{cc}
e^{-\imath \phi /2} & 0 \\ 0 & e^{\imath \phi /2}
\end{array}\right).
\label{eq:trans-mat}
\end{align}

The $Nth$-order intensity moments of polarization mode $\hat{b}_1$, $\langle\hat{b}_1^{\dagger N} \hat{b}_1^{N}\rangle_k$, abbreviated as $B_k$, can be measured while the LC and HWP are set to $\phi_k$ and $\theta_k$ respectively, and the measurement is iterated for $(N+1)^2$ different angle pairs (meaning $k=1,2,...,(N+1)^2$). Using Eqs. (\ref{eq:coord-trans}) and (\ref{eq:trans-mat}), we can relate the measurement of $B_k$ to all the $N$th-order coherences forming a set of $(N+1)^2$ linear equations:
\begin{align}
B_k = R_k^{m n} \rho_{m n} \label{eq:set-eqs}
\end{align}
where m and n are by convention summed from $0$ to $N$, and $R_k^{m n} = \binom{N}{n} \binom{N}{m} e^{\imath \phi_k (m-n)} \times\\ (\sin \theta_k + \cos \theta_k)^{n+m}(\sin \theta_k - \cos \theta_k)^{2N-n-m} /2^N$. \\
The linear set of equations in Eq. (\ref{eq:set-eqs}) has a solution from which $\rho_{m n}$ can be uniquely determined when the rank of $R_k^{m n}$ equals $(N+1)^2$. This condition can be met by choosing a set of $(N+1)^2$ pairs ${(\theta_k,\phi_k)}$ , where $k=1,...,(N+1)^2$ and for each such pair the $N$-fold coincidence $B_k$ is measured. An Example for possible sets are given for different $N$ in Table \ref{tab:table1}.

\begin{table}
\caption{\label{tab:table1} Angles of HWP rotation and LC phase in QST measurements for a specific $N$. Each QST for a fixed number of photons, $N$, constitute a set of consecutive $N$-fold coincidence measurements in polarization mode $b_1$, where each measurement in this set is done while the HWP and LC are prepared in a pair combination of $\theta$ and $\phi$ from this table. Each QST measurement set is composed out of all $(N+1)^2$ pair combination in this table (except for odd $N$).}
\begin{ruledtabular}
\begin{tabular}{lcc}
$N$ & $\theta$ & $\phi$ \\
\hline

2 &
$-\frac{\pi}{3},0,\frac{\pi}{3}$ &
$0,\frac{2\pi}{3},\frac{4\pi}{3}$ \\

3\footnotemark[1] &
$-\frac{7\pi}{18},0,\frac{7\pi}{18}$ &
$\frac{\pi}{5},\frac{3\pi}{5},\pi,\frac{7\pi}{5},\frac{9\pi}{5}$ \\

4 &
$-\frac{13\pi}{30},-\frac{3\pi}{20},\frac{3\pi}{20},\frac{13\pi}{30}$ &
$\frac{\pi}{5},\frac{3\pi}{5},\pi,\frac{7\pi}{5},\frac{9\pi}{5}$ \\

5\footnotemark[1] &
$-\frac{13\pi}{30},-\frac{2\pi}{15},0,\frac{2\pi}{15},\frac{13\pi}{30}$ &
$\frac{2\pi}{7},\frac{4\pi}{7},\frac{6\pi}{7},\frac{8\pi}{7},\frac{10\pi}{7},\frac{12\pi}{7},2\pi $ \\
\end{tabular}
\end{ruledtabular}
\footnotemark[1]{In the case of odd $N$, a measurement for the pair $\theta = -\frac{\pi}{4},\phi = 0$ should be added to all other $(N+1)^2-1$ pair combinations from the table.}
\end{table}

Determination of a quantum state is implemented by reconstruction of the density matrix from a set of measurements and the set of linear equations in Eq. (\ref{eq:set-eqs}) relating the measurements to the coherences. A naive way to perform this reconstruction is by forming $\rho$ and $B$ as vectors, and $R$ as a $(N+1)^2 \times (N+1)^2$ matrix with columns indexed by $mn$ and rows by $k$ and inverting Eq. (\ref{eq:set-eqs}) to find the density matrix: $\rho_{mn} = (R_k^{mn})^{-1} B_k$. This can be problematic as the resulting density matrix can be non-physical due to experimental noise, for example the estimated density matrix may well not generally be semi-positive definite. The most probable physical density matrix is obtained by employing the method of maximum-likelihood;
this estimation technique is a well known method in statistics that is used in QST to determine the most probable density matrix compliant with a set of measurements \cite{QBIT01PRA}.

$\mathcal{L}(\rho_{est})$, defined as
\begin{align} \label{eq:likelihood}
\mathcal{L}(\rho_{est}) = \sum_{k = 1}^{(N+1)^2}\frac{\left(R_k^{m n}\rho_{est,mn} - B_k \right)^2}{B_k}.
\end{align}
is a convex probability function that an estimated density matrix $\rho_{est}$ would produce the set of measurements of N-fold coincidence, $B_k$.

This function was checked using a Monte-Carlo (MC) method to be a consistent, unbiased and efficient likelihood function estimator of the density matrix (for different input states, with and without simulated noise). The problem of quantum state estimation for a given N is now reduced to finding the minimum for the function $\mathcal{L}(\rho_{est})$ using an optimization algorithm.
It was noted before \cite{KOSUT04ArXiv} that the likelihood function is convex and hence its global minimum can be found by a convex optimization algorithm. To perform convex optimization we used the free optimization package YALMIP \cite{YALMIP10} under the physical constraints on the density matrix of unit trace, hermiticity, and being positive semi-definite.

\emph{Two photon Overlap }--- In the analysis of the previous section it was assumed that the photons are indistinguishable in all degrees of freedom (except for polarization). The actual experiment, however, involves the interference of a CS and SPDC which in general might not have the same spatial and spectral modes. In the experiment we use a single mode fiber to ensure spatial overlap and a band-pass filter for enhanced spectral overlap. To verify the validity of the indistinguishability assumption we measure the two-photon overlap between the two sources (defined below), which is the lowest order of interference possible since the lowest non-vacuum contribution of the SPDC has two photons. Recently, heralded single photons from SPDC were shown \cite{JIN11PRA} to overlap a single photon from CS with high visibility in a Hong-Ou-Mandel interference \cite{HOM87PRL}. An overlap of unity implies perfect indistinguishability between the photons while an overlap of $0$ implies complete distinguishability. We note that in the case of highly distinguishable photons the recently introduced method of hidden differences \cite{ADAMSON07PRL} may be used.

Considering the overall quantum state of a system fed by two separate beams, SPDC and a CS,
\begin{align} \label{eq:instate}
|\Psi\rangle = |\Psi_{SPDC}\rangle \otimes |\Psi_{CS}\rangle
\end{align}
where the quantum states of SPDC and CS can be written in terms of their two-photon spectral wave-functions $\Phi(\omega_1,\omega_2)$ and $A(\omega_1)A(\omega_2)$ respectively as
\begin{align} \label{eq:instate2}
|\Psi_{SPDC}\rangle = \frac{1}{\sqrt{U}}{\int}{\int} d\omega_1 d\omega_2 \Phi(\omega_1,\omega_2) |\omega_1\rangle|\omega_2\rangle \\
|\Psi_{CS}\rangle =  \frac{1}{\sqrt{V}}{\int}{\int} d\omega_1 d\omega_2 A(\omega_1)A(\omega_2) |\omega_1\rangle|\omega_2\rangle
\end{align}
where $U$ and $V$ are proportional to the $2$-fold coincidence of the SPDC and CS respectively and serve as normalization factors. We superpose the two states on two input ports of a beam splitter. The $2$-fold coincidence in one of the output ports of the beam splitter, denoted by $P^{(2)}_{b_1}$, will be proportional to
\begin{widetext}
\begin{eqnarray}\label{eq:overlap}
\nonumber P^{(2)}_{b_1}(\varphi_{CS}) &=& {\int}{\int} d\omega_1 d\omega_2 |\Phi (\omega_1,\omega_2)|^2 +
{\int}{\int} d\omega_1 d\omega_2 |A(\omega_1)A(\omega_2)|^2 +  \\
\nonumber & & {\int}{\int} d\omega_1 d\omega_2 d\omega^{\prime}_1 d\omega^{\prime}_2 (e^{\imath 2\varphi_{CS}} A(\omega_1)A(\omega_2) \Phi^{\ast}(\omega^{\prime}_1,\omega^{\prime}_2)
 + e^{-\imath 2\varphi_{CS}} A^{\ast}(\omega_1)A^{\ast}(\omega_2) \Phi(\omega^{\prime}_1,\omega^{\prime}_2)) \\
  & &\ \ \ \ \ \ \ \ \ \ \ \ \ \ \ = U + V + 2 \sqrt{U V} \cos(2\varphi_{CS}) | \langle\Psi_{SPDC} | \Psi_{CS}\rangle| .
\end{eqnarray}
\end{widetext}
where $\varphi_{CS}$ is the CS phase (relative to the SPDC).\\
From Eq.(\ref{eq:overlap}) we can see that the last term is an oscillating cosine function of $\varphi_{CS}$, while the other two terms contribute a constant background. Therefore, by measuring $P^{(2)}_{b_1}$, $U$, and $V$, and scanning the phase $\varphi_{CS}$, we can fit $P^{(2)}_{b_1}$ to a cosine function with a constant background and extract the \emph{overlap} of the normalized SPDC and CS states, $| \langle\Psi_{SPDC} |\Psi_{CS}\rangle|$, which can result any value between $0$ (orthogonal states) and $1$ (identical states).\\ \\
\section{Experimental Results}

The experimental setup consists three consecutive parts: state preparation, state transformation, and detection. NOON states are prepared for different values of N by setting pair amplitude ratio between the CS and SPDC inputs denoted by $\gamma$, and the phase of CS relative to the SPDC represented by $\varphi_{CS}$ \cite{AFEKSCIENCE} at the $\pm 45^{\circ}$ $\hat{X}, \hat{Y}$ polarizations basis respectively (see Fig.\ref{fig:setup}). A transformation on the prepared states is performed using a phase shifting LC and a rotated HWP to $(N+1)^2$ different pairs of angles $(\phi,\theta)$, which are given in Table \ref{tab:table1}. The resulting states are measured using five single photon counting modules (SPDC) in polarization mode $\hat{b}_1$. An overall polarization extinction ratio in the setup was measured to be better than $1/250$ to ensure that no polarization effect degraded the tomographic analysis.

We have also performed a theoretical simulation of the experiment using the definitions of a CS and SPDC \cite{IntroQuantOPt}. The simulation accounts for all sources of loss in the experiment in the abstract loss model \cite{achilles2004photon} using a single parameter $\eta$ representing the overall transmission of the setup. The simulation also assumes that all the photons arrive in a single identical spatial and temporal mode. Good agreement of the simulation and the experiment implies that the assumption of indistinguishability is valid.

\begin{figure}[t!!]
\includegraphics[width=0.48\textwidth]{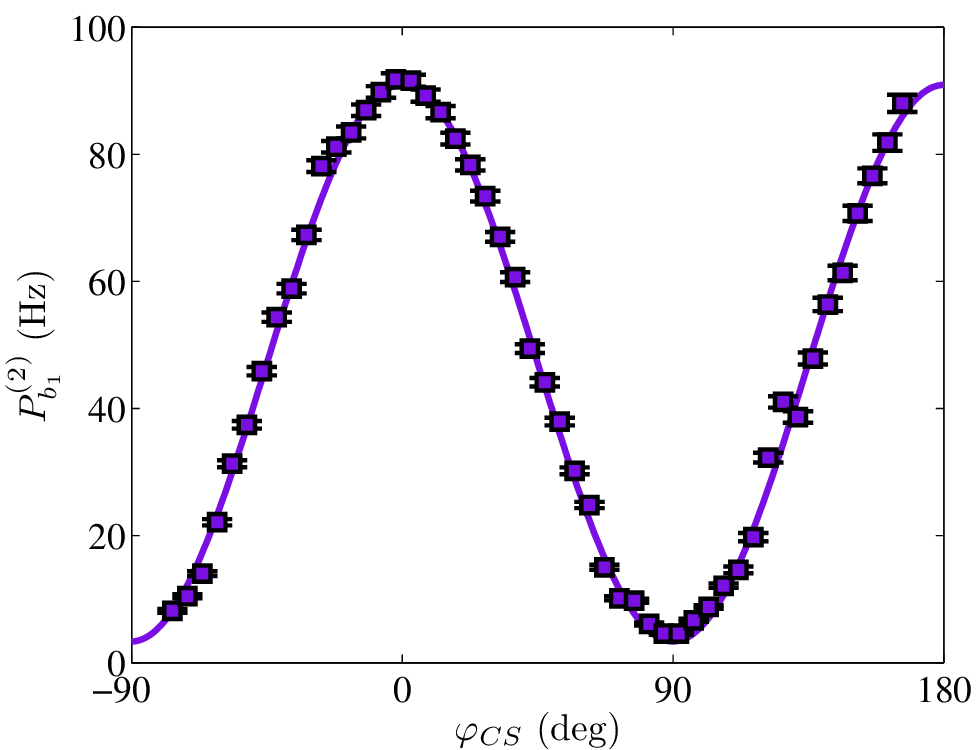}
\caption{\label{fig:overlapfig} (Color online) Two photon overlap measurement of CS and SPDC. $2$-fold coincidence measurement of an output port of a beam splitter, $P_{b1}^{(2)}$, fed by CS in one of its input port and SPDC in the other. Error bars indicate $\pm\sigma$ statistical uncertainty. Solid line is obtained by fitting the data according to Eq. (\ref{eq:overlap}) resulting with an overlap of $|\langle\Psi_{SPDC} |\Psi_{CS}\rangle| = 0.970\pm0.020$ (with $95\%$ confidence bound, see text).}
\end{figure}

We start by measuring the two photon overlap (TPO) function (defined in the previous section) which describes the two-photon interference of the CS with the SPDC on a beam splitter. To perform this, in the actual experiment, we use our setup (see Fig. \ref{fig:setup}), while $\theta = 0, \phi = \pi/2$ serves as a beam splitter mixing the SPDC and CS, and measure the 2-fold coincidence in polarization mode $\hat{b}_1$, one of its output ports. The LC in the CS path controls the phase, $\varphi_{CS}$, which enables the TPO fringes. Fig. \ref{fig:overlapfig} shows the measurement of the two photon overlap,
 for SPDC and CS 2-fold coincidence rate of about $50 Hz$, which is low enough to significantly neglect
 the next contributing order, which arises from a three photon event with one photon loss.
As follows from Eq. (\ref{eq:overlap}) by fitting the measurement to a cosine function of $\varphi_{CS}$ with a
 constant background (2-fold coincidence rate of SPDC and CS) the TPO between the CS
 and the SPDC is extracted, $|\langle\Psi_{SPDC} |\Psi_{CS}\rangle| = 0.970\pm0.020$ (with $95\%$ confidence bound).

\begin{figure}[t!!]
\includegraphics[width=0.48\textwidth]{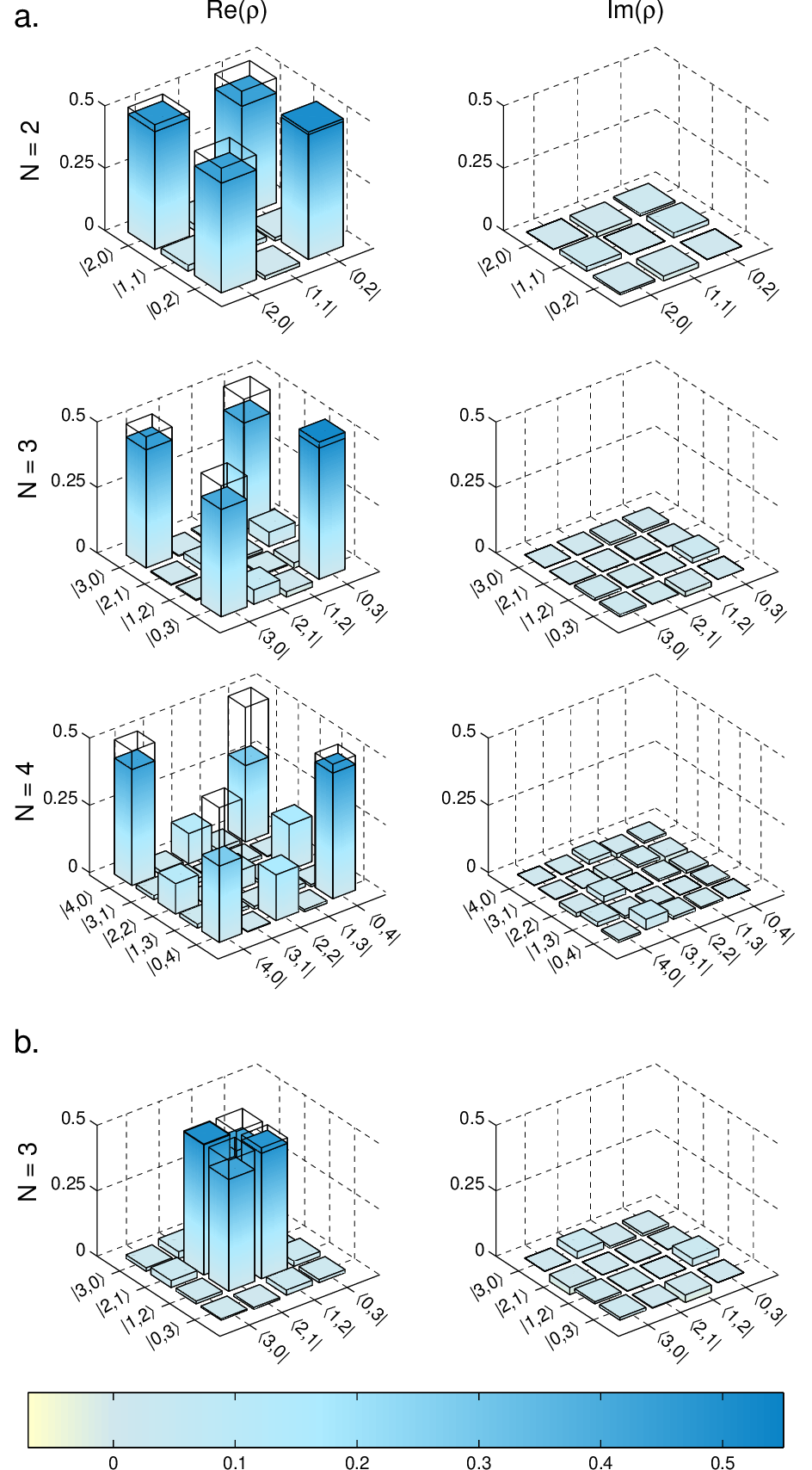}
\caption{\label{fig:NOON_DM} (Color online) Experimental results: real and imaginary parts of density matrices
 of (\texttt{a}) NOON states for $N = 2,3,4$ (\texttt{b}) correlated photon holes of 3,0 and 0,3 (in colored bars), reconstructed in the photon-number basis of two modes states labeled
 $|m,n\rangle$, where m and n are the number of photons in polarization modes $\hat{a}_1$ and $\hat{a}_2$
 respectively (see Fig. \ref{fig:setup}). Solid frames represent density matrices of the ideal states.}
\end{figure}

The experimental reconstructed density matrices for the prepared states for $N = 2,3,4$ are shown in Fig. \ref{fig:NOON_DM}a.
The fidelity parameter $\mathcal{F}$ of an estimated state $\rho_{est}$ to overlap a theoretical state $\rho_{th}$ is defined as
\begin{align}
\mathcal{F}(\rho_{est},\rho_{th}) = \left( Tr \sqrt{ \sqrt{\rho_{th}} \rho_{est} \sqrt{\rho_{th}} } \right)^2.  \label{eq:fidelity}
\end{align}

It gives a quantitative measure of the correspondence between theoretical and experimental density matrices of the states. We compare the experimental results both to the theoretical NOON states, and to the theoretical simulation of our setup. The experimental error in the fidelity is calculated from a MC simulation of the statistical error of the $N$-fold coincidence rate and applying the density matrix reconstruction procedure. The fidelities of the density matrices of NOON state for $N = 2, 3, 4$ (Fig. \ref{fig:NOON_DM}a) to the theoretical ideal NOON states are $0.931\pm0.004$, $0.895\pm0.021$ and $0.721\pm0.063$ respectively, while their fidelities to the states resulting from the simulation are $0.984\pm0.003$, $0.986\pm0.014$ and $0.930\pm0.089$ respectively with $\pm\sigma$ statistical uncertainty. The high fidelity of the experiment to the simulation indicates that the main source of deviation from an ideal NOON state indeed stems from the low efficiency of our setup and not from photon distinguishability of the two photon state.

\begin{figure}[t!!]
\includegraphics[width=0.48\textwidth]{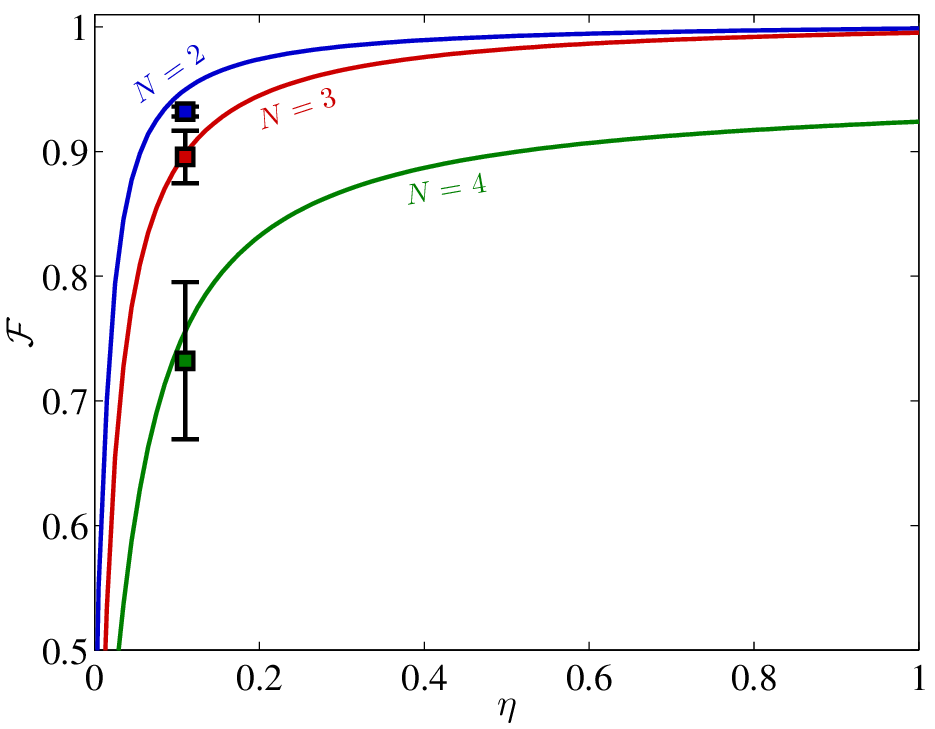}
\caption{\label{fig:FidelityVsEta} (Color online) NOON states fidelity dependence on setup transmission $\eta$. Fidelity of the states generated by our setup to ideal NOON states are shown as a function of the setup transmission $\eta$ for $N = 2, 3, 4$ in blue, red and green respectively. Simulations (solid lines) are shown together with the measurements (squares) in the current setup transmission, $\eta = 0.11$. Error bars indicate $\pm\sigma$ statistical uncertainty.}
\end{figure}

Fig. \ref{fig:FidelityVsEta} shows the dependence of the fidelity (of NOON states to ideal NOON states)
 on $\eta$ (defined as the coincidence to singles ratio of the SPDC, being the overall setup transmission),
 for N = 2, 3, 4. A simulation of our system (as described above) for different values of $\eta$ with our
 procedure of density matrix reconstruction was employed. The low fidelity for lower transmission shows the susceptibility of the high-NOON states to loss. The fidelity in our setup is shown here to degrade mostly by the setup transmission. Our current setup transmission,
 $\eta = 0.11\pm0.02$, is determined experimentally by the coincidences to singles ratio of the SPDC; it is mainly limited by three major technicalities: detectors efficiency, coupling of
 light into the single-mode fiber, and the band-pass filter transmission.

Additionally, we show in Fig. \ref{fig:NOON_DM}b density matrix reconstruction for yet another class of quantum states of light, the CPH states \cite{CPH10PRL}. These states exhibit absence of multiphoton coincidence events between two modes amid a constant background. We show here as an example the state reconstruction of a CPH with holes in the states $|3,0\rangle_{A,B}$ and $|0,3\rangle_{A,B}$, where $A$ and $B$ correspond in our setup (Fig. \ref{fig:setup}) to polarization modes $\hat{a}_1$ and $\hat{a}_2$ respectively. This is achieved by setting $\gamma=3$ and $\varphi_{CS}=\pi$. The fidelity of the measured state to the ideal CPH state is $0.888\pm0.037$.

\section{Conclusion}
We have performed a complete analysis of High-NOON states generated by mixing CS with SPDC. We reconstructed density matrices for the measured states showing high fidelities with respect to other methods \cite{SHALM09NAT,ADAMSON07PRL} in a relatively simple setup with no background substraction. The quantum state tomography was performed under the simplifying assumption that the photons are indistinguishable
 in all degrees of freedom but polarization. We checked this assumption by measuring the
 two photon overlap. We analyzed our results by calculating the fidelity of the measured states with ideal NOON states, and with states resulting from a simulation of our setup. The fidelity of the measured states with ideal NOON states is limited in our configuration by photon loss, which is dominated by three elements: the single-mode fiber, the band-pass filter, and the single-photon detectors. Roughly, the first two contribute an amount of 50\% loss each, and the detection efficiency of each single photon detector is about 50\%. We thus stress the need in biphoton sources which can be spectrally and spatially matched to CS and "true" photon-number-resolving detectors, or single photon detectors with high quantum efficiencies.

\begin{acknowledgments}
I. A. gratefully acknowledges the support of the Ilan
Ramon Grant. Financial support of this research by the ERC grant QUAMI, the Minerva Foundation and the Crown Photonics Center is gratefully acknowledged.
Correspondence and requests for materials
should be addressed to Y. I. ~(email: yonatan.israel@weizmann.ac.il).
\end{acknowledgments}

\bibliography{text}

\end{document}